\documentclass[aps,pra,twocolumn,showpacs,superscriptaddress,letterpaper]{revtex4-1}
\usepackage{graphicx,amsmath,amssymb}
\usepackage[utf8x]{inputenc}
\usepackage{url}
\usepackage{float}
\usepackage{siunitx}
\usepackage{color}
\usepackage{xspace}

\newcommand{\ie}{i.e.\@\xspace}
\newcommand{\eg}{e.g.\@\xspace}

\renewcommand{\bm}[1]{\boldsymbol{\mathbf{#1}}}
\newcommand{\bra}{\left\langle}
\newcommand{\ket}{\right\rangle}
\newcommand{\im}{\operatorname{Im}}
\newcommand{\re}{\operatorname{Re}}

\begin{document}

\title{Multiple scattering of polarized light in disordered media exhibiting short-range structural correlations}

\author{Kevin Vynck}
\email{kevin.vynck@institutoptique.fr}
\affiliation{LP2N, CNRS - Institut d'Optique Graduate School - Univ. Bordeaux, F-33400 Talence, France}
\author{Romain Pierrat}
\author{R\'{e}mi Carminati}
\affiliation{ESPCI Paris, PSL Research University, CNRS, Institut Langevin, 1 rue Jussieu, F-75005, Paris, France}

\date{\today}

\begin{abstract}
   We develop a model based on a multiple scattering theory to describe the diffusion of polarized light in disordered
   media exhibiting short-range structural correlations. Starting from exact expressions of the average field and the
   field spatial correlation function, we derive a radiative transfer equation for the polarization-resolved specific intensity that is
   valid for weak disorder and we solve it analytically in the diffusion limit. A decomposition of the specific intensity
   in terms of polarization eigenmodes reveals how structural correlations, represented via the standard anisotropic
   scattering parameter $g$, affect the diffusion of polarized light. More specifically, we find that propagation
   through each polarization eigenchannel is described by its own transport mean free path that depends on $g$ in a
   specific and non-trivial way.
\end{abstract}

\maketitle

\section{Introduction}\label{sec:1}

Electromagnetic waves propagating in disordered media are progressively scrambled by refractive index fluctuations and,
thanks to interference, result into mesoscopic phenomena, such as speckle correlations and weak
localization~\cite{Akkermans2011, Sheng2010}. Polarization is an essential characteristic of electromagnetic waves that,
considering the ubiquity of scattering processes in science, prompted the development of research in statistical
optics~\cite{Goodman2015, Brosseau1998} and impacted many applications, from optical imaging in biological
tissues~\cite{Tuchin2006} to material spectroscopy (\eg, rough surfaces)~\cite{Maradudin2007}, and radiation transport
in turbulent atmospheres~\cite{Andrews2005, Shirai2003}. Although the topic has experienced numerous developments and
outcomes in the past decades, recent studies have revealed that much remains to be explored and understood on the
relation between the microscopic structure of scattering media and the polarization properties of the scattered field.
In particular, it was found that important information about the morphology of a disordered medium is contained in
the three-dimensional (3D) polarized speckles produced in the near-field above its surface~\cite{Apostol2003,
Carminati2010, Parigi2016} and in the spontaneous emission properties of a light source in the bulk~\cite{Caze2010,Sapienza2011a}.
Similarly, the light scattered by random ensembles of large spheres was shown to exhibit unusual polarization features
due to the interplay between the various multipolar scatterer resonances~\cite{Schmidt2015}.

The fact that light transport is affected by the microscopic structural properties of disordered media is well known.
Structural correlations, coming from the finite scatterer size or from the specific morphology of porous
materials~\cite{Torquato2005, RojasOchoa2004, Garcia2007}, typically translate into an anisotropic phase function,
$p(\cos \theta)$, which describes the angular response of a single scattering event with the scattering angle $\theta$. The
average cosine of the phase function, known as the anisotropic scattering factor, $g=\bra \cos \theta \ket$ (with
$-1 \leq g \leq 1$), then leads to the standard definition of the transport mean free path (the average distance after
which the direction of light propagation is completely randomized) as $\ell^*=\ell/(1-g)$, where $\ell$ is the
scattering mean free path (the average distance between two scattering events). Single scattering anisotropy naturally
affects how the polarization diffuses in disordered media, one of the most notable findings being that circularly
polarized light propagates on longer distances compared to linearly polarized light in disordered media exhibiting
forward single scattering ($g>0$) ---the so-called ``circular polarization memory effect''~\cite{MacKintosh1989a,
Xu2005a, Gorodnichev2007}.

Recent observations in mesoscopic optics also motivate deeper investigations on polarized light transport in correlated
disordered media. Indeed, numerical simulations revealed that uncorrelated ensembles of point scatterers cannot exhibit
3D Anderson localization due to the vector nature of light~\cite{Skipetrov2014, Bellando2014}. By contrast, it was found that the
interplay between short-range structural correlations and scatterer resonances could yield the opening of a 3D photonic
gap in disordered systems~\cite{Edagawa2008, Liew2011} and promote localization phenomena at its
edges~\cite{Imagawa2010}. To date, the respective role of polarization and structural correlations on mesoscopic optical
phenomena remains largely to be clarified.

Theoretically describing the propagation of polarized light in disordered media exhibiting structural correlations is a
difficult task. A first approach consists in using the vector radiative transfer equation~\cite{Chandrasekhar1960,
Papanicolaou1975, Mishchenko2006}, in which electromagnetic waves are described via the Stokes parameters and the
scattering and absorption processes are related via energy conservation arguments. The various incident polarizations
(linear, circular) and the single scattering anisotropy are explicitly implemented, thereby allowing for the
investigation of a wide range of problems~\cite{Amic1997, Gorodnichev2014}. A second approach relies on a transfer
matrix formalism based on a scattering sequence picture, where each scattering event (possibly anisotropic) yields a partial
redistribution of the light polarization along various directions~\cite{Akkermans1988, Xu2005, Rojas-Ochoa2004a}. The
approach is phenomenological, yet very intuitive, making it possible to gain important physical insight into mesoscopic
phenomena such as coherent backscattering~\cite{Akkermans1988}.

The most \textit{ab-initio} approach to wave propagation and mesoscopic phenomena in disordered systems is the so-called
multiple scattering theory, which directly stems from Maxwell's equations and relies on perturbative expansions on the
scattering potential~\cite{Sheng2010, Akkermans2011}. The formalism is often used to investigate mesoscopic
phenomena, such as short and long-range (field and intensity) correlations or coherent backscattering, in a large
variety of complex (linear or nonlinear) media, including disordered dielectrics and atomic clouds. Unfortunately, it
also rapidly gains in complexity when the vector nature of light is considered. In fact, multiple scattering theory for
polarized light has so far been restricted to uncorrelated disordered media only~\cite{Stephen1986, MacKintosh1988,
Ozrin1992, VanTiggelen1996, VanTiggelen1999, Muller2002, Vynck2014}.

In this article, we present a model based on multiple scattering theory that describes how the diffusion of polarized
light is affected by short-range structural correlations, thereby generalizing previous models limited to uncorrelated
disorder. We do not aim at developing a complete theory for polarization-related mesoscopic phenomena in correlated
disordered media but at showing that, by a series of well-controlled approximations, important steps towards this
objective can be made. Starting from the (exact) Dyson and the Bethe-Salpeter equations for the average field and the
field correlation function, we derive a radiative transfer equation for the polarization-resolved specific
intensity in the limit of short-range structural correlations and weak scattering. To analyze the impact of short-range
structural correlations on the diffusion of polarization, we then apply a $P_1$ approximation and decompose the
polarization-resolved energy density into ``polarization eigenmodes'', as was done previously for uncorrelated
disordered media~\cite{Ozrin1992, Muller2002, Vynck2014}. An interesting outcome of this decomposition is the
observation that each polarization eigenmode is affected independently and differently by short-range structural
correlations. More precisely, each mode is characterized by a specific transport mean free path, and thus a specific
attenuation length (describing the depolarization process) for its intensity. The transport mean free path of each
eigenmode depends non-trivially on the anisotropy factor $g$, and differently from the $(1-g)^{-1}$ rescaling
well known for the diffusion of scalar waves.

The paper is organized as follows. The radiative transfer equation for polarized light is derived
\textit{ab-initio} in Sect.~\ref{sec:2}. The diffusion limit and the eigenmode decomposition are applied in
Sect.~\ref{sec:3}. In Sect.~\ref{sec:4}, we discuss the model and the results deduced from it, paying special attention to 
the consistency of the approximations that have been made. Our conclusions are given in Sect.~\ref{sec:5}. 
Technical details about the average Green's function, the range of validity of the short-range structural correlation approximation, 
and the particular case of uncorrelated disorder, are presented in Appendices~\ref{sec:A1}--\ref{sec:A3}, respectively.

\section{Radiative transfer for polarized light}\label{sec:2}

\subsection{Spatial field correlation}

We consider a disordered medium described by a real dielectric function of the form
$\epsilon(\bm{r})=1+\delta\epsilon(\bm{r})$, where $\delta\epsilon(\bm{r})$ is the fluctuating part with the
statistical properties
\begin{equation}\label{eq:disorder}
   \bra \delta\epsilon(\bm{r}) \ket = 0,
   \qquad \bra \delta\epsilon(\bm{r}) \delta\epsilon(\bm{r}') \ket = u f(\bm{r}-\bm{r}')
\end{equation}
where $\bra\ldots\ket$ indicates ensemble averaging. The function
$f(\bm{r}-\bm{r}')$ describes the structural correlation of the medium and $u$ is an amplitude whose expression will be
derived below. We assume that the medium is statistically isotropic and invariant by translation. Considering a
monochromatic wave with free-space wavevector $k_0=\omega/c=2\pi/\lambda$, $\omega$ being the frequency,
$\lambda$ the wavelength and $c$ the speed of light in vacuum, the electric field $\textbf{E}$ satisfies the vector 
propagation equation
\begin{equation}
   \nabla \times \nabla \times \bm{E}(\bm{r})-k_0^2 \epsilon(\bm{r}) \bm{E}(\bm{r})=i \mu_0 \omega \bm{j}(\bm{r}),
\end{equation}
where the current density $\bm{j}(\bm{r})$ describes a source distribution in the disordered medium. 
Introducting the dyadic Green's function $G_{ik}$, the $i$th component of the electric field reads
\begin{equation}\label{eq:Efield-green}
E_i(\bm{r})=i \mu_0  \omega \int G_{ik}(\bm{r},\bm{r}') j_k(\bm{r}') d\bm{r}',
\end{equation} 
where implicit summation of repeated indices is assumed.
The spatial correlation function of the electric field $\bra E_i(\bm{r}) E_j^\star(\bm{r}') \ket$ obeys the Bethe-Salpeter equation
\begin{multline}\label{eq:BS_field}
   \bra E_i(\bm{r}) E_j^\star(\bm{r}') \ket = \bra E_i(\bm{r}) \ket \bra E_j^\star (\bm{r}') \ket
\\
   + k_0^4 \int \bra G_{im}(\bm{r}-\bm{r}_1) \ket \bra G_{jn}^\star(\bm{r}'-\bm{r}_1') \ket
\\
   \times \Gamma_{mnrs} (\bm{r}_1,\bm{r}_1',\bm{r}_2,\bm{r}_2') \bra E_r (\bm{r}_2) E_s^\star (\bm{r}_2') \ket d\bm{r}_1 d\bm{r}_1' d\bm{r}_2 d\bm{r}_2'
\end{multline}
that can be derived from diagrammatic calculations~\cite{Akkermans2011,Sheng2010}. In this expression
the superscript $\star$ denotes complex conjugation, and
$\Gamma_{mnrs}$ is the four-point irreducible vertex that describes all possible scattering sequences
between four points. In Eq.~(\ref{eq:BS_field}), the first term in the right-hand side corresponds to the ballistic 
intensity, that is attenuated due to scattering at the scale of the scattering mean free path $\ell$, and
the second term describes the multiple-scattering process. Note that at this level, Eq.~(\ref{eq:BS_field}) 
is an exact closed-form equation.

It is also interesting to remark that the field correlation function $\bra E_i(\bm{r}) E_j^\star(\bm{r}') \ket$ is
one of the key quantities in statistical optics (where it is usually denoted by cross-spectral density matrix), since it
encompasses the polarization and coherence properties of fluctuating fields in the frequency domain~\cite{Goodman2015, Brosseau1998}. 
The study of light fluctuations in 3D multiple scattering media has stimulated a revisiting of the concepts of degree of polarization 
and coherence~\cite{Setala2002, Dennis2007, Refregier2014, Gil2014,Dogariu2015}, initially defined for 2D paraxial fields.

To proceed further, we assume weak disorder, such that the scattering mean free path $\ell$
 is much larger than the wavelength ($k_0\ell \gg 1$). In this regime, only the two diagrams
for which the field and its complex conjugate follow the same trajectories (the so-called ladder and most-crossed
diagrams) contribute to the average intensity. The ladder diagrams are the root of radiative transport theory, that describes
the transport of intensity as an incoherent process. The most-crossed diagrams are responsible for weak localization and
coherent backscattering. In the ladder approximation and assuming independent scattering, the four-point irreducible
vertex reduces to
\begin{multline}
   \Gamma_{mnrs}(\bm{r}_1,\bm{r}_1',\bm{r}_2,\bm{r}_2')
\\
   \begin{split}
      & = \bra \delta\epsilon(\bm{r}_1) \delta\epsilon(\bm{r}_1') \ket \delta(\bm{r}_1-\bm{r}_2) \delta(\bm{r}_1'-\bm{r}_2') \delta_{mr} \delta_{ns}
   \\
      & = u f(\bm{r}_1-\bm{r}_1') \delta(\bm{r}_1-\bm{r}_2) \delta(\bm{r}_1'-\bm{r}_2') \delta_{mr} \delta_{ns},
   \end{split}
\end{multline}
yielding
\begin{multline}\label{eq:BS_field2}
   \bra E_i(\bm{r}) E_j^\star(\bm{r}') \ket = \bra E_i(\bm{r}) \ket \bra E_j^\star (\bm{r}') \ket
\\
   + uk_0^4 \int \bra G_{im}(\bm{r}-\bm{r}_1) \ket \bra G_{jn}^\star(\bm{r}'-\bm{r}_1') \ket
\\
   \times f(\bm{r}_1-\bm{r}_1') \bra E_m (\bm{r}_1) E_n^\star (\bm{r}_1') \ket d\bm{r}_1 d\bm{r}_1'.
\end{multline}
We consider the source to be a point electric dipole located at $\bm{r}_0$, such that
\begin{equation}
   j_k(\bm{r})=-i\omega p_k \delta(\bm{r}-\bm{r}_0),
\end{equation}
where $p_k$ is the dipole moment along direction $k$. 
Equation~(\ref{eq:Efield-green}) simplifies into $E_i(\bm{r})= \mu_0 \omega^2 \, G_{ik}(\bm{r}-\bm{r}_0) p_k$ and the
Bethe-Salpeter equation (\ref{eq:BS_field2}) can be rewritten in terms of the dyadic Green's function in the form
\begin{multline}\label{eq:BS_green}
   \bra G_{ik}(\bm{r}-\bm{r}_0) G_{jl}^\star(\bm{r}'-\bm{r}_0) \ket = \bra G_{ik}(\bm{r}-\bm{r}_0) \ket \bra G_{jl}^\star (\bm{r}'-\bm{r}_0) \ket
\\
   + uk_0^4 \int \bra G_{im}(\bm{r}-\bm{r}_1) \ket \bra G_{jn}^\star(\bm{r}'-\bm{r}_1') \ket f(\bm{r}_1-\bm{r}_1')
\\
   \times \bra G_{mk} (\bm{r}_1-\bm{r_0}) G_{nl}^\star (\bm{r}_1'-\bm{r}_0) \ket d\bm{r}_1 d\bm{r}_1'.
\end{multline}
Using the change of variables $\bm{r}-\bm{r}_0=\bm{R}+\bm{X}/2$ and $\bm{r}'-\bm{r}_0=\bm{R}-\bm{X}/2$, and 
transforming Eq.~(\ref{eq:BS_green}) into reciprocal space, with $\bm{K}$ and $\bm{q}$ the reciprocal variables of
$\bm{R}$ and $\bm{X}$ respectively, we finally obtain
\begin{widetext}
   \begin{multline}\label{eq:BS_green_fourier}
      \bra G_{ik}\left(\bm{q}+\frac{\bm{K}}{2}\right) G_{jl}^\star\left(\bm{q}-\frac{\bm{K}}{2}\right) \ket
         = \bra G_{ik}\left(\bm{q}+\frac{\bm{K}}{2}\right) \ket \bra G_{jl}^\star\left(\bm{q}-\frac{\bm{K}}{2}\right) \ket
      + uk_0^4 \bra G_{im}\left(\bm{q}+\frac{\bm{K}}{2}\right) \ket \bra G_{jn}^\star\left(\bm{q}-\frac{\bm{K}}{2}\right) \ket
   \\
      \times \int f(\bm{q}-\bm{q}') \bra G_{mk} \left(\bm{q}'+\frac{\bm{K}}{2}\right) G_{nl}^\star
         \left(\bm{q}'-\frac{\bm{K}}{2}\right) \ket \frac{d\bm{q}'}{8\pi^3}.
   \end{multline}
\end{widetext}
A direct resolution of Eq.~(\ref{eq:BS_green_fourier}) is possible for $f(\bm{q}-\bm{q}')=1$,
and this approach was used in Ref.~\cite{Vynck2014} to study the coherence and polarization properties of light in
an uncorrelated disordered medium. In the case of a medium with structural correlations, a direct resolution is out of reach
and we need to follow a different strategy.

\subsection{From field correlation to radiative transfer}

In this section we derive a radiative transfer equation for polarized light.
We proceed by evaluating the average Green's tensor $\bra \bm{G} \ket$, that obeys the Dyson
equation~\cite{Akkermans2011}. In its most general form, it reads~\cite{Tai1993}
\begin{equation}\label{eq:averageG}
   \bra \bm{G} (\bm{q}) \ket = \left[ k_0^2 \bm{I} - q^2 \bm{P}(\hat{\bm{q}}) - \bm{\Sigma}(\bm{q}) \right]^{-1},
\end{equation}
with $\bm{I}$ the unit tensor, $\bm{P}(\hat{\bm{q}})=\bm{I}-\hat{\bm{q}} \otimes \hat{\bm{q}}$ the transverse
projection operator, $\hat{\bm{q}}=\bm{q}/q$ and $q=|\bm{q}|$. $\bm{\Sigma}(\bm{q})$ is the self-energy, which contains
the sum over all multiple scattering events that cannot be factorized in the averaging process. As shown in
Appendix~\ref{sec:A1}, for arbitrary structural correlations, $\bm{\Sigma}(\bm{q})$ is non-scalar. 
The problem can be simplified by assuming short-range structural correlations,
in which case $\bm{\Sigma}(\bm{q})=\Sigma(\bm{q})\bm{I}$.  The average Green's tensor can then be written as
\begin{equation}\label{eq:Green-mulet}
   \bra \bm{G} (\bm{q}) \ket = \bra G(\bm{q}) \ket \left( \bm{I} - \frac{\bm{q} \otimes \bm{q}}{k_0^2-\Sigma(\bm{q})} \right),
\end{equation}
with $\bra G(\bm{q}) \ket =[k_0^2-q^2-\Sigma(\bm{q})]^{-1}$ the scalar Green's function. In a dilute medium,
the scattering events are assumed to take place on large distances compared to the wavelength (near-field interactions between 
scatterers can be neglected). In this case, the average Green's tensor $\bra \bm{G} (\bm{q}) \ket$ can be reduced to its transverse
component~\cite{Arnoldus2003}, yielding
\begin{equation}
   \bra \bm{G} (\bm{q}) \ket \simeq \bra G(\bm{q}) \ket \bm{P}(\hat{\bm{q}}).
\end{equation}
After some simple algebra, the first term in the right-hand side in Eq.~(\ref{eq:BS_green_fourier}) can be written as
\begin{multline}
   \bra G_{ik}\left(\bm{q}+\frac{\bm{K}}{2}\right) \ket \bra G_{jl}^\star\left(\bm{q}-\frac{\bm{K}}{2}\right) \ket
\\
   = M_{ik} M'_{jl}\frac{\bra G(\bm{q}+\bm{K}/2) \ket - \bra G^\star(\bm{q}-\bm{K}/2) \ket}
      {2 \bm{q} \cdot \bm{K} + \Sigma(\bm{q}+\bm{K}/2) - \Sigma^\star(\bm{q}-\bm{K}/2)},
\end{multline}
where we have defined the polarization factors $M_{ik}=\delta_{ik} - (q_i + K_i/2) (q_k + K_k/2)/|\bm{q}+\bm{K}/2|^2$ and
$M'_{jl}=\delta_{jl} - (q_j - K_j/2) (q_l - K_l/2)/|\bm{q}-\bm{K}/2|^2$. In a dilute medium, we can assume that $|\bm{K}| \ll
|\bm{q}|$. This means that there are two different space scales in the correlation function of Green's tensor: A short scale associated to $\bm{q}$ and corresponding to the dependence on direction of the specific intensity that we will introduce in Eq.~(\ref{eq:specific_intensity}), and a large scale associated to $\bm{K}$ and corresponding to the dependence of the specific intensity on position. This leads to
\begin{multline}\label{eq:Green1}
   \bra G_{ik}\left(\bm{q}+\frac{\bm{K}}{2}\right) \ket \bra G_{jl}^\star\left(\bm{q}-\frac{\bm{K}}{2}\right) \ket
\\
   = (\delta_{ik} - \hat{q}_i \hat{q}_k) (\delta_{jl} - \hat{q}_j \hat{q}_l)
      \frac{\bra G(\bm{q}) \ket - \bra G^\star(\bm{q}) \ket}{2 \bm{q} \cdot \bm{K} + 2i \im[\Sigma(\bm{q})]}.
\end{multline}
The self-energy $\Sigma(\bm{q})$ renormalizes the propagation constant in the medium by defining a complex effective
permittivity $\epsilon_\text{eff}=1-\Sigma(\bm{q})/k_0^2$. The real part of $\Sigma$ yields a change in the phase velocity,
and the imaginary an attenuation of the field amplitude due to scattering. Hence, we can write
\begin{equation}\label{eq:avGreen}
   \bra G(\bm{q}) \ket = \frac{1}{k_0^2 \re[\epsilon_\text{eff}] - q^2 + i k_0^2 \im[\epsilon_\text{eff}]}.
\end{equation}
Since $\im[\epsilon_\text{eff}] \ll \re[\epsilon_\text{eff}]$ in a dilute medium, we can rewrite
Eq.~(\ref{eq:avGreen}) using the identity
\begin{equation}
   \lim_{\varepsilon \rightarrow 0} \frac{1}{x-x_0-i \varepsilon} = \operatorname{PV}\left[ \frac{1}{x-x_0} \right] - i \pi \delta (x-x_0),
\end{equation}
where $\operatorname{PV}$ stands for principal value.
Defining $q_e=k_0\sqrt{\re[\epsilon_\text{eff}]}$ as an effective wavevector, Eq.~(\ref{eq:Green1}) becomes
\begin{multline}\label{eq:Green2}
   \bra G_{ik}\left(\bm{q}+\frac{\bm{K}}{2}\right) \ket \bra G_{jl}^\star\left(\bm{q}-\frac{\bm{K}}{2}\right) \ket
\\
   = (\delta_{ik} - \hat{q}_i \hat{q}_k) (\delta_{jl} - \hat{q}_j \hat{q}_l) \frac{\pi \delta(q_e^2-q^2)}{ i \bm{q} \cdot \bm{K} - \im[\Sigma(\bm{q})]}.
\end{multline}
In order to derive a radiative transfer equation, we then introduce the quantity $L_{ijkl}$  by the relation
\begin{multline}\label{eq:specific_intensity}
   \bra G_{ik}\left(\bm{q}+\frac{\bm{K}}{2}\right) G_{jl}^\star\left(\bm{q}-\frac{\bm{K}}{2}\right) \ket
\\
      = \frac{4\pi^2}{q_e} \delta(q_e^2 - q^2) L_{ijkl}(\bm{K},q_e \hat{\bm{q}}).
\end{multline}
Here, we assume that the correlation function of Green's tensor propagates on shell, {\it i.e.} with a wavevector $q=q_e$. The impact of the on-shell approximation, which is the key step to solve the Bethe-Salpeter equation in the presence of  structural correlations, will be discussed in Sec.~\ref{sec:4}. From Eqs.~(\ref{eq:Green2}) and (\ref{eq:specific_intensity}), we can rewrite the Bethe-Salpeter equation (\ref{eq:BS_green_fourier}) in the form
\begin{multline}
   \frac{4\pi^2}{q_e} \delta(q_e^2 - q^2) L_{ijkl}(\bm{K},q_e \hat{\bm{q}})
\\
   = \frac{\pi \delta(q_e^2 - q^2)}{ i \bm{q} \cdot \bm{K} - \im[\Sigma(\bm{q})]}
      \left[\vphantom{\int} (\delta_{ik} - \hat{q}_i \hat{q}_k) (\delta_{jl} - \hat{q}_j \hat{q}_l)\right.
\\
   + u k_0^4 (\delta_{im} - \hat{q}_i \hat{q}_m) (\delta_{jn} - \hat{q}_j \hat{q}_n)
\\
   \left. \times  \frac{4\pi^2}{q_e} \int f(\bm{q}-\bm{q}') \delta(q_e^2 - q'^2) L_{mnkl}(\bm{K},q_e \hat{\bm{q}}')
   \frac{d\bm{q}'}{8\pi^3} \right].
\end{multline}
Integrating both sides of the equation over $q$, performing the integral on the right-hand side over $q'$, and using the relation $\int_0^\infty f(\bm{r}) \delta(r^2-r_0^2) r^2 dr=r_0f(\bm{r}=r_0 \hat{\bm{r}})/2$, we obtain
\begin{multline}\label{eq:preRTE}
   L_{ijkl}(\bm{K},q_e \hat{\bm{q}})
\\
   =\frac{q_e}{4\pi}\frac{1}{ i q_e \hat{\bm{q}} \cdot \bm{K} - \im[\Sigma(q_e \hat{\bm{q}})]}
      \left[\vphantom{\int} (\delta_{ik} - \hat{q}_i \hat{q}_k) (\delta_{jl} - \hat{q}_j \hat{q}_l) \right.
\\
   + \frac{u k_0^4}{4\pi} (\delta_{im} - \hat{q}_i \hat{q}_m) (\delta_{jn} - \hat{q}_j \hat{q}_n)
\\
   \left.\times \int f(q_e (\hat{\bm{q}}-\hat{\bm{q}}')) L_{mnkl}(\bm{K}, q_e \hat{\bm{q}}') d\hat{\bm{q}}' \right].
\end{multline}
The quantity $L_{ijkl}(\bm{K},q_e \hat{\bm{q}})$ is proportional to the specific intensity introduced in radiative transfer theory~\cite{Chandrasekhar1960}, and has the meaning of a local and directional radiative flux. Actually, Eq.~(\ref{eq:preRTE}) can be cast in the form of a radiative transfer equation, as we will now show.

Since the disordered medium is statistically isotropic and translational-invariant, the correlation function $f$ only depends on $|\hat{\bm{q}}-\hat{\bm{q}}'|$, or equivalently on $\hat{\bm{q}} \cdot \hat{\bm{q}}'$. It is directly related to the classical phase function $p(\hat{\bm{q}}\cdot\hat{\bm{q}}')$ of radiative transfer theory as
\begin{equation}\label{eq:correlation-phase}
   f(q_e |\hat{\bm{q}}-\hat{\bm{q}}'|) = A \, p(\hat{\bm{q}}\cdot\hat{\bm{q}}'),
\end{equation}
where $A$ is a constant whose value is determined by energy conservation, and $\int p(\hat{\bm{q}}\cdot\hat{\bm{q}}') d\hat{\bm{q}} = 4\pi$.
To order $(k_0\ell)^{-1}$ and for short-range structural correlations, one has $\im[\Sigma(q_e \hat{\bm{q}})] =
-q_e/\ell$ and $u=6\pi/k_0^4\ell$ (these results are derived in Appendix~\ref{sec:A1}). This allows us to rewrite Eq.~(\ref{eq:preRTE})
in its final form 
\begin{widetext}
   \begin{equation}\label{eq:RTE}
      \left[ i \hat{\bm{q}} \cdot \bm{K} + \frac{1}{\ell} \right] L_{ijkl}(\bm{K},\hat{\bm{q}})
         = \frac{1}{4\pi} (\delta_{ik} - \hat{q}_i \hat{q}_k) (\delta_{jl} - \hat{q}_j \hat{q}_l)
         + \frac{3 A}{8 \pi \ell} (\delta_{im} - \hat{q}_i \hat{q}_m) (\delta_{jn} - \hat{q}_j \hat{q}_n)
         \int p(\hat{\bm{q}} \cdot \hat{\bm{q}}') L_{mnkl}(\bm{K}, \hat{\bm{q}}') d\hat{\bm{q}}'
   \end{equation}
\end{widetext}
where an implicit summation over $m$ and $n$ is assumed.
This expression takes the form of a radiative transfer equation (RTE) for the polarization-resolved specific intensity.
It differs from the standard vector radiative transfer equation~\cite{Chandrasekhar1960} in the sense that it is not written
in terms of Stokes vector, but using a fourth-order tensor representing the specific intensity for polarized light, and relating two
input and two output polarization components. Nevertheless, the various terms in Eq.~(\ref{eq:RTE}) have a very clear
physical meaning. The first and second terms on the left-hand-side respectively describe the total variation of specific
intensity along direction $\hat{\bm{q}}$ and the extinction of the ballistic light due to scattering (\ie, Beer-Lambert's
law). The first and second terms on the right-hand-side describe the increase of specific intensity along direction
$\hat{\bm{q}}$ due to the presence of a source, and to the light originally propagating along direction $\hat{\bm{q}}'$
and being scattered along $\hat{\bm{q}}$, respectively.

Conservation of energy requires the scattering losses to be compensated by the gain due to scattering after
integration over all angles. The energy conservation relation has to be written on the intensity, \ie by setting $i=j$ and
summing over polarization components in Eq.~(\ref{eq:RTE}), in the form
\begin{multline}
   \frac{1}{\ell}\sum_i \int L_{iikl}(\bm{K},\hat{\bm{q}}) d\hat{\bm{q}}  = \frac{3 A}{8\pi\ell}
\\
   \times \sum_{i,m} \int (\delta_{im} - \hat{q}_i \hat{q}_m)^2 p(\hat{\bm{q}} \cdot \hat{\bm{q}}') L_{mmkl}(\bm{K}, \hat{\bm{q}}') d\hat{\bm{q}}' d\hat{\bm{q}}.
\end{multline}
This leads to the following relation on the coefficient $A$
\begin{equation}\label{eq:energy_conservation_constant}
   \frac{3}{8\pi} \sum_m \int (\delta_{im} - \hat{q}_i \hat{q}_m)^2 p(\hat{\bm{q}} \cdot \hat{\bm{q}}') d\hat{\bm{q}} = \frac{1_i}{A},
\end{equation}
where $1_i$ is the unit vector. At this stage, we have obtained a transport equation for polarized light
[Eq.~(\ref{eq:RTE})] that takes the form of a RTE. This equation stems directly from the Dyson and Bethe-Salpeter equations, 
fulfills energy conservation, and is valid for dilute media and short-range correlated disorder.

\section{Diffusion of polarization}\label{sec:3}

\subsection{$P_1$ approximation}

In short-range correlated media, the phase function $p(\hat{\bm{q}} \cdot \hat{\bm{q}}')$ is expected to be
quasi-isotropic. It can therefore be expanded into a Legendre series, which, to order $\hat{\bm{q}} \cdot \hat{\bm{q}}'$,
reads
\begin{equation}\label{eq:phasefunction}
   p(\hat{\bm{q}} \cdot \hat{\bm{q}}') = 1 + 3g (\hat{\bm{q}} \cdot \hat{\bm{q}}'),
\end{equation}
where $g$ is the anisotropic scattering factor, defined as
\begin{equation}
   g =\frac{1}{4\pi} \int p({\bm{q}} \cdot \hat{\bm{q}}') \hat{\bm{q}} \cdot \hat{\bm{q}}' d\hat{\bm{q}},
\end{equation}
and satisfying
\begin{equation}
   g \hat{\bm{q}}' =\frac{1}{4\pi} \int p({\bm{q}} \cdot \hat{\bm{q}}') \hat{\bm{q}} d\hat{\bm{q}}.
 \end{equation}
Inserting Eq.~(\ref{eq:phasefunction}) into Eq.~(\ref{eq:RTE}), the RTE can be rewritten as
\begin{multline}\label{eq:newRTE}
   \left[ i \hat{\bm{q}} \cdot \bm{K} + \frac{1}{\ell} \right] L_{ijkl}(\bm{K},\hat{\bm{q}})
   = \frac{1}{4\pi} (\delta_{ik} - \hat{q}_i \hat{q}_k) (\delta_{jl} - \hat{q}_j \hat{q}_l)
\\
   + \frac{3 A}{2 \ell} (\delta_{im} - \hat{q}_i \hat{q}_m) (\delta_{jn} - \hat{q}_j \hat{q}_n)
   \\
   \times \left[
      L^{(0)}_{mnkl}(\bm{K})+ \frac{3g}{4 \pi } \bm{j}_{mnkl}(\bm{K}) \cdot \hat{\bm{q}}\right],
\end{multline}
where $L^{(0)}_{ijkl}$ and $\bm{j}_{ijkl}$ are the (polarization-resolved)
irradiance and radiative flux vector, respectively, defined as
\begin{align}
   L^{(0)}_{ijkl}(\bm{K}) & = \frac{1}{4\pi} \int L_{ijkl}(\bm{K},\hat{\bm{q}}) d\hat{\bm{q}},
\\
   \bm{j}_{ijkl}(\bm{K}) & = \int \hat{\bm{q}} L_{ijkl}(\bm{K},\hat{\bm{q}}) d\hat{\bm{q}}.
\end{align}

To gain insight into the effect of short-range correlations on the propagation of polarized light, it is convenient to
investigate the diffusion limit, which is reached after propagation on distances much larger than the 
scattering mean free path $\ell$. In this limit, the specific intensity becomes quasi-isotropic. 
Expanding $L_{ijkl}$ into Legendre polynomials $P_n$ to first order in $\hat{\bm{q}}$, we have 
\begin{equation}\label{eq:P1_approx}
   L_{ijkl}(\bm{K},\hat{\bm{q}}) = L^{(0)}_{ijkl}(\bm{K}) + \frac{3}{4\pi} \bm{j}_{ijkl}(\bm{K}) \cdot \hat{\bm{q}}
\end{equation}
which is the so-called $P_1$ approximation.
Inserting Eq.~(\ref{eq:P1_approx}) into Eq.~(\ref{eq:newRTE}) and calculating the zeroth and first moments of the
resulting equation (which amounts to performing the integrations $\int - d\hat{\bm{q}}$ and $\int - \hat{\bm{q}} d\hat{\bm{q}}$,
respectively), we eventually arrive to a pair of equations relating $L^{(0)}_{ijkl}$ and $\bm{j}_{ijkl}$:
\begin{widetext}
   \begin{align}
      i \bm{K} \cdot \bm{j}_{ijkl}(\bm{K}) + \frac{4\pi}{\ell} L^{(0)}_{ijkl}(\bm{K})
         & = \frac{2}{3} S_{ijkl} + \frac{4\pi}{\ell} A S_{ijmn} L^{(0)}_{mnkl}(\bm{K}), \label{eq:zeroth_moment}
   \\
      -\frac{4\pi}{3} K^2 \ell L^{(0)}_{ijkl}(\bm{K}) + i \bm{K} \cdot \bm{j}_{ijkl}(\bm{K})
         & = i g \frac{9A}{8\pi} \int (\delta_{im} - \hat{q}_i \hat{q}_m) (\delta_{jn} - \hat{q}_j \hat{q}_n)
         \left( \bm{j}_{mnkl}(\bm{K}) \cdot \hat{\bm{q}} \right) \left( \bm{K} \cdot \hat{\bm{q}} \right)
         d\hat{\bm{q}}.\label{eq:first_moment}
   \end{align}
\end{widetext}
Here, we have defined
\begin{equation}
   S_{ijkl}=\frac{3}{8\pi} \int (\delta_{ik} - \hat{q}_i \hat{q}_k) (\delta_{jl} - \hat{q}_j \hat{q}_l) d\hat{\bm{q}},
\end{equation}
and used the relations $\int (\delta_{im} - \hat{q}_i \hat{q}_m) (\delta_{jn} - \hat{q}_j \hat{q}_n) \hat{\bm{q}}
d\hat{\bm{q}}=0$, $\int \hat{q}_i \hat{q}_j d\hat{\bm{q}}= 4\pi / 3\delta_{ij}$ and $\int \hat{q}_i \hat{q}_j
\hat{q}_k d\hat{\bm{q}}= 0$. The additional complexity of the polarization mixing due to structural correlations can be
apprehended from Eq.~(\ref{eq:first_moment}), where the relation between $L^{(0)}_{ijkl}$ and $\bm{j}_{ijkl}$ in terms
of input and output polarization components becomes particularly intricate as soon as $g \neq 0$. Much deeper insight
into the diffusion of polarized light can be gained via an eigenmode decomposition, as shown below.

\subsection{Polarization eigenmodes}

Analytical expressions for all terms in the $L_{ijkl}^{(0)}(\bm{K})$ and $\bm{j}_{ijkl}(\bm{K})$ tensors can be
obtained by solving Eqs.~(\ref{eq:zeroth_moment}) and (\ref{eq:first_moment}), which we have done imposing $\bm{K}$ to
be along one of the main spatial directions, without loss of generality, and using the software Mathematica~\cite{Mathematica}. The
obtained expressions at this stage are long and complicated, containing in particular high-order terms in powers of $K$
and $g$ (that are not physical and will be neglected below). We now introduce a polarization-resolved energy density
$U_{ijkl}=6\pi / cL_{ijkl}^{(0)}$ and decompose it in terms of ``polarization eigenmodes''
as in Refs.~\cite{Ozrin1992, Muller2002, Vynck2014}:
\begin{equation}\label{eq:eigenmode_decomposition}
   U_{ijkl}(\bm{K}) = \sum_p U^{(p)}(\bm{K}) \left|ij\ket_p \bra kl \right|_p.
\end{equation}
The eigenvalues $U^{(p)}$ provide the characteristic length and time scales of the diffusion of each eigenmode and the projectors
$\left|ij\ket_p \bra kl \right|_p$, which will be denoted by ``polarization eigenchannels'', relate input polarization pairs
$(k,l)$ to output polarization pairs $(i,j)$. The $U_{ijkl}$ is represented as a $9 \times 9$ matrix (9 pairs of
polarization components in input and output) and is diagonalized using Mathematica, leading again to full analytical expressions.

At this stage, the obtained expressions still depend on the coefficient $A$, originally defined in
Eq.~(\ref{eq:correlation-phase}) and used to ensure energy conservation in the RTE, Eq.~(\ref{eq:RTE}). To predict how
$A$ depends on structural correlations, we rely on the particular case of the Henyey-Greenstein (HG) phase function~\cite{Henyey1941}
\begin{equation}\label{eq:HG-phasefunction}
   p_\text{HG}(\hat{\bm{q}} \cdot \hat{\bm{q}}') = \frac{1-g}{\left[1+g^2-2g (\hat{\bm{q}} \cdot \hat{\bm{q}}')\right]^{3/2}}.
\end{equation}
The HG phase function is very convenient since it provides a closed-form expression with $g$ as a single parameter, and
approximates the phase functions of a wide range of disordered media (\eg, interstellar dust clouds, biological
tissues). The energy conservation equation, Eq.~(\ref{eq:energy_conservation_constant}), can be solved analytically in
this case, yielding the surprisingly simple relation
\begin{equation}\label{eq:A_HG}
   A_\text{HG} = \frac{1}{1+g^2/2}.
\end{equation}
Note that the modification in energy conservation due to structural correlations appears at order $g^2$.

We can finally insert Eq.~(\ref{eq:A_HG}) into the eigenvalues and eigenvectors found from
Eq.~(\ref{eq:eigenmode_decomposition}) and develop analytical expressions valid to orders $K^2$ (diffusion
approximation) and $g$ (weakly correlated disorder). The eigenvectors take the expressions already obtained for
uncorrelated disorder~\cite{Ozrin1992, Muller2002, Vynck2014}
\begin{align}\label{eq:eigenvectors}
   \left|kl\ket_{1} & = \frac{1}{\sqrt{3}} \delta_{kl}, \nonumber \\
   \left|kl\ket_{2,3,4} & = \frac{1}{\sqrt{2}} (\delta_{ka} \delta_{lb} - \delta_{kb} \delta_{la}), \nonumber \\
   \left|kl\ket_{5} & = \frac{1}{\sqrt{2}} (\delta_{ka} \delta_{la} - \delta_{kb} \delta_{lb}), \nonumber \\
   \left|kl\ket_{6,7,8} & = \frac{1}{\sqrt{2}} (\delta_{ka} \delta_{lb} + \delta_{kb} \delta_{la}), \nonumber \\
   \left|kl\ket_{9} & = \frac{1}{\sqrt{6}} (\delta_{ka} \delta_{la} + \delta_{kb} \delta_{lb}- 2 \delta_{kc} \delta_{lc}).
\end{align}
The first eigenchannel is the scalar mode, relating uniformly pairs of identical polarization components ($xx$, $yy$ and
$zz$), which describe the classical intensity, between themselves. The other eigenchannels either redistribute
nonuniformly the energy between pairs of identical polarization ($p=5$ and $9$), thereby participating as well in the
propagation of the classical intensity, or are concerned with pairs of orthogonal polarizations ($xy$, $xz$, etc), which
can participate, for instance, in magneto-optical media in which light polarization can rotate~\cite{MacKintosh1988,
VanTiggelen1996, VanTiggelen1999}.

The eigenvalues take the form of the solution of the diffusion equation in reciprocal space
\begin{equation}\label{eq:diffusionsolution}
   U^{(p)}(\bm{K}) = \frac{1}{\mathcal{D}^{(p)} K^2 + \mu_a^{(p)} c},
\end{equation}
where $\mathcal{D}^{(p)}$ and $\mu_a^{(p)}$ are the diffusion constant and attenuation coefficient of the $p$th
polarization mode. The eigenmode energy densities in real space therefore read
\begin{equation}
   U^{(p)}(\bm{R}) = \frac{1}{4\pi \mathcal{D}^{(p)} R} \exp\left[- \frac{R}{\ell_\text{eff}^{(p)}} \right],
\end{equation}
with $R=|\bm{R}|$ and $\ell_\text{eff}^{(p)}=\sqrt{\mathcal{D}^{(p)}/\mu_a^{(p)} c}$, which is an effective attenuation
length, describing the depolarization process.

Table~\ref{tab:diffcorr} summarizes the diffusion constants, attenuation coefficients and effective attenuation lengths
of the different polarization eigenchannels. As in the case of uncorrelated
disorder previously studied in Ref.~\cite{Vynck2014}, all modes exhibit different diffusion constants, thereby spreading at different speeds, and
only the scalar mode persists at large distances ($\ell_\text{eff}^{(1)}=\infty$), all other modes being attenuated on a
length scale on the order of a mean free path.

\begin{table*}
   \caption{Summary of the diffusion constants $\mathcal{D}^{(p)}$, attenuation coefficients $\mu_a^{(p)}$ and effective
   attenuation lengths $\ell_\text{eff}^{(p)}$ characterizing the diffusion properties of the energy density through the
   individual polarization eigenchannels and the depolarization process. Note that all quantities are given to order
   $g$. Quite remarkably, structural correlations, via the scattering asymmetry factor $g$, are found to affect
   differently and independently each mode.}\label{tab:diffcorr}
   \begin{ruledtabular}
      \def\arraystretch{1.5}
      \begin{tabular}{c|c|c|c|c|c|c}
         $p$ & 1 & 2 & 3,4 & 5,6 & 7,8 & 9 \\ \hline
         $\mathcal{D}^{(p)}$ & $ \left(1-g\right)^{-1} \frac{c\ell}{3}$ & $\left(\frac{1}{2}-\frac{9}{20}g\right)^{-1} \frac{c\ell}{3}$ & $\left(\frac{1}{2}-\frac{3}{20}g\right)^{-1} \frac{c\ell}{3}$ & $\left(\frac{7}{10}-\frac{69}{100}g\right)^{-1} \frac{c\ell}{3}$ & $\left(\frac{7}{10}-\frac{39}{100}g\right)^{-1} \frac{c\ell}{3}$ & $\left(\frac{7}{10}-\frac{29}{100}g\right)^{-1} \frac{c\ell}{3}$ \\ \hline
         $\mu_a^{(p)}$ & $0$ & $\frac{1}{\ell}$ & $\frac{1}{\ell}$ & $\frac{3}{7\ell}$ & $\frac{3}{7\ell}$ & $\frac{3}{7\ell}$ \\ \hline
         $\ell_\text{eff}^{(p)}$ & $\infty$ & $\left( 1-\frac{9}{20}g \right)^{-1} \sqrt{\frac{2}{3}}\ell$ & $\left( 1-\frac{3}{20}g \right)^{-1} \sqrt{\frac{2}{3}}\ell$ & $\left( 1-\frac{69}{140}g \right)^{-1} \frac{\sqrt{10}}{3}\ell$ & $\left( 1-\frac{39}{140}g \right)^{-1} \frac{\sqrt{10}}{3}\ell$ & $\left( 1-\frac{29}{140}g \right)^{-1} \frac{\sqrt{10}}{3}\ell$
      \end{tabular}
   \end{ruledtabular}
\end{table*}

\begin{figure}
   \centering
	\includegraphics[width=0.5\textwidth]{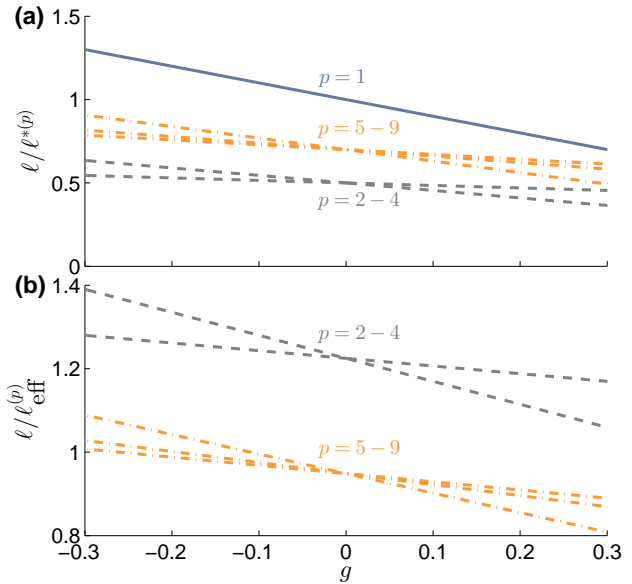}
   \caption{(Color online only) Evolution of (a) the
   transport coefficient, $1/\ell^{*(p)}$, and (b) the attenuation coefficient, $1/\ell_\text{eff}^{(p)}$, of
   polarization eigenmodes with short-range structural correlations. The coefficients are given in units of $1/\ell$ and
   shown on a restricted range of $g$ since the model is expected to remain valid to first order near $g=0$. The scalar
   mode ($p=1$, cyan solid curve) has a transport coefficient scaling as $(1-g)$ and an attenuation coefficient equal to
   zero (not shown). The polarization modes $p=2$--$4$ (gray dashed curves) and $p=5$--$9$ (orange dot-dashed curves)
   exhibit different slopes, indicating that both their transport properties are affected differently by short-range
   structural correlations.}
   \label{fig1}
\end{figure}

More interestingly, our study brings new information on the influence of short-range structural correlations on transport and depolarization.
Let us first remark that we properly recover the diffusion constant of the scalar mode, $\mathcal{D}=c\ell^*/3$ with $\ell^*=\ell/(1-g)$ the
transport mean free path, which is a good indication of the validity of the model. The second and
more interesting finding in this study is the fact that the propagation characteristics of each polarization mode is
affected independently and differently by short-range structural correlations. One may have anticipated that the
diffusion constant of each polarization mode would be simply rescaled by the $(1-g)^{-1}$ factor relating scattering and
transport mean free paths. Instead, we show that a transport mean free path can be defined for each polarization mode,
$\ell^{*(p)}=3\mathcal{D}^{(p)}/c$ and its dependence on the anisotropy factor $g$ can change significantly, as shown in Fig.~\ref{fig1}(a). This, in
turn, implies that the spatial attenuation of each polarization mode (due to depolarization) is affected differently by
structural correlations, as shown in Fig.~\ref{fig1}(b).

\section{Discussion}\label{sec:4}

Previous studies based on the multiple scattering theory for the propagation of polarized light relied on the
\textit{direct} resolution of the Bethe-Salpeter equation, Eq.~(\ref{eq:BS_green_fourier}), using an expansion of the average Green's
tensors and its correlation function to order $K^2$ (diffusion approximation). This strategy is however
possible only for uncorrelated disorder, for which $f(\bm{q}-\bm{q}')=1$. Here, we proposed an alternative strategy based on
the derivation of a transport equation taking the form of an RTE, which allowed us to reach the same final goal (eigenmode decomposition) including
short-range structural correlations. This strategy, however, involves an additionnal approximation that has some implications. To
clarify this point, let us consider our predictions in the limit of an uncorrelated disorder. Setting $g=0$ in the
predictions of Table~\ref{tab:diffcorr} yields the values reported in Table~\ref{tab:diffuncorr}. An alternative
straightforward derivation from Eqs.~(\ref{eq:zeroth_moment}) and (\ref{eq:first_moment}), which yields the same
results, is proposed in Appendix~\ref{sec:A3}. Compared to previous results (see, \eg, Ref.~\onlinecite{Vynck2014}), we
observe that the eigenvectors, or polarization eigenchannels, remain unchanged, but the eigenvalues are now 1, 3 and
5-fold degenerate, yielding the same attenuation coefficients $\mu_a^{(p)}$ but different diffusion constants
$\mathcal{D}^{(p)}$. This apparent discrepancy can be explained by the on-shell approximation, which ``smoothes out''
the polarization dependence in the correlation function of Green's tensor. Nevertheless, it is important to note that the
\textit{average} diffusion constants for the various degenerate modes are strictly identical:
\begin{equation}
\frac{1}{3} \left(\frac{6}{5}c\ell + 2\frac{2}{5}c\ell \right) = 2 \frac{c\ell}{3},
\end{equation}
and
\begin{equation}
\frac{1}{5} \left(2\frac{230}{343}c\ell + 2\frac{130}{343}c\ell + \frac{290}{1029}c\ell \right) = \frac{10}{7}\frac{c\ell}{3}
\end{equation}
This brings us to the conclusion that the model is consistent with the approximations that have been made.

\begin{table}[H]
   \caption{Summary of the diffusion constants $\mathcal{D}^{(p)}$, attenuation coefficients $\mu_a^{(p)}$ and effective
   attenuation lengths $\ell_\text{eff}^{(p)}$ characterizing the diffusion properties of the energy density through the
   individual polarization eigenchannels for an uncorrelated disorder ($g=0$).}\label{tab:diffuncorr}
   \begin{ruledtabular}
      \def\arraystretch{1.5}
      \begin{tabular}{l|l|l|l}
         $p$ & 1 & 2-4 & 5-9 \\ \hline
         $\mathcal{D}^{(p)}$ & $\frac{c\ell}{3}$ & $2 \frac{c\ell}{3}$ & $\frac{10}{7}\frac{c\ell}{3}$ \\ \hline
         $\mu_a^{(p)}$ & $0$ & $\frac{1}{\ell}$ & $\frac{3}{7\ell}$ \\ \hline
         $\ell_\text{eff}^{(p)}$ & $\infty$ & $\sqrt{\frac{2}{3}}\ell$ & $\frac{\sqrt{10}}{3}\ell$
      \end{tabular}
   \end{ruledtabular}
\end{table}

A second point deserving a comment is the fact that the attenuation length $1/\mu_a^{(p)}$ of the polarization eigenmodes
does not depend on $g$ to first order, the effect of short-range structural correlations on the spatial
decay of polarization away from the source being implemented via the definition of mode-specific transport mean free
paths. This picture contrasts with previous studies based on the phenomenological transfer matrix approach~\cite{Xu2005,
Rojas-Ochoa2004a}, which relate the depolarization length $\ell_p$ for linearly polarized light to the \textit{scalar}
transport mean free path via a linear relation with $g$. In this sense, our model provides a different perspective on
this basic problem of light transport in disordered media. Intuitively, this picture also appears more physically sound,
since it is known that the relation between depolarization and transport mean free path varies with the incident
polarization (linear, circular) or in presence of magneto-optical effects~\cite{MacKintosh1988, VanTiggelen1996}.

Related to this point, it is also important to discuss the validity of the diffusion limit to retrieve depolarization
coefficients. Reaching the regime of diffusive transport typically requires light to experience several multiple
scattering events. However, as pointed out previously (see, \eg, Ref.~\onlinecite{Gorodnichev2014}), this limit can hardly be
achieved for the polarization modes, for which the depolarization occurs on the scale of a mean free path. It is then
legitimate to question the accuracy of the expressions reported in Table~\ref{tab:diffcorr}. Nevertheless,
we do not expect this question to impact our claim that different polarization modes are individually and differently
affected by short-range structural correlations. Actually, the established RTE for the polarization-resolved specific
intensity, Eq.~(\ref{eq:RTE}), like the standard vector radiative transfer equation, does not assume diffusive
transport. On this aspect, our study constitutes a very good starting point to investigate the validity of the diffusion
approximation, which may be done either numerically by solving the RTE by Monte-Carlo methods, or analytically by adding
higher-order Legendre polynomials $P_n$ in the following steps.

Finally, let us remark that the results of our model, in which disorder is described by a continuous and randomly fluctuating function of position [Eq.~(\ref{eq:disorder})], should apply not only to heterogeneous materials with complex textit{connected} morphologies (e.g., porous media) but also to random ensembles of finite-size scatterers. Indeed, the Fourier transform of the structural correlation $f(\bm{r}-\bm{r}')$ directly leads to the definition of the phase function $p(\hat{\bm{q}}\cdot\hat{\bm{q}}')$ [Eq.~(\ref{eq:correlation-phase})], which is the same function to which one arrives when investigating light scattering by finite-size scatterers (it is, in this case, defined from the differential scattering cross-section). For the sake of broadness of applications and convenience, the final results here have been given for the HG phase function [Eq.~(\ref{eq:HG-phasefunction})] but other phase functions (e.g., Mie for spherical scatterers) may be used to describe specific disordered media. Note that for ensembles of finite-size scatterers, the short-range correlation approximation restricts the validity range of the model to small scatterers.

\section{Conclusion}\label{sec:5}

To conclude, we have proposed a model based on multiple scattering theory to describe the propagation of polarized light
in disordered media exhibiting short-range structural correlations. Our results assume weak disorder ($k_0\ell \gg 1$),
short-range structural correlations (first order in $g$), and are obtained in the ladder approximation. Starting from the exact
Dyson and Bethe-Salpeter equations for the average field and the field correlation, we have derived a RTE for the
polarization-resolved specific intensity [Eq.~(\ref{eq:RTE})] and applied the $P_1$ approximation to investigate the
propagation of polarized light in the diffusion limit. Interestingly, we have found that the polarization modes, described
so far for uncorrelated disorder only, are independently and differently affected by short-range structural
correlations. In practice, each mode is described by its own transport mean free path, which does not trivially depend
on $g$ (see Table~\ref{tab:diffcorr}). 

In essence, our study partly unveils the intricate relation between the complex morphology of disordered media and the
polarization properties of the scattered intensity. The road towards a possible description of polarization-related
mesoscopic phenomena in correlated disorder is long, yet we hope that the present work, which highlights several
theoretical challenges when dealing with polarized light and structural correlations, will motivate future
investigations. The model may be generalized, for instance, by including the most-crossed diagrams in the derivation to
enable the study of phenomena such as weak localization, or frequency dependence to investigate ---via a generalized
RTE--- the temporal response to incident light pulses. Another line of research could be to study the impact of
short-range structural correlations on spatial coherence properties, which appears extremely relevant to the optical
characterization of complex nanostructured media~\cite{Dogariu2015}.

\section*{Acknowledgements}

The authors acknowledge John Schotland for stimulating discussions. This work is supported by LABEX WIFI (Laboratory of
Excellence within the French Program ``Investments for the Future'') under references ANR-10-LABX-24 and
ANR-10-IDEX-0001-02 PSL$^*$, by INSIS-CNRS via the LILAS project and the CNRS ``Mission for Interdisciplinarity''
via the NanoCG project.

\appendix

\section{Average Green's tensor}\label{sec:A1}

The average Green's tensor $\bra \bm{G}\ket$ describes the propagation of the average field in the disordered medium and
is related to the free-space Green's tensor $\bm{G}_0$ via the Dyson equation~\cite{Akkermans2011, Sheng2010}
\begin{equation}
   \bra \bm{G}(\bm{q}) \ket = \bm{G}_0(\bm{q}) + \bm{G}_0(\bm{q}) \bm{\Sigma}(\bm{q}) \bra \bm{G}(\bm{q}) \ket,
\end{equation}
where $\bm{\Sigma}$ is the self-energy, which contains the sums over all multiply scattered events that cannot be
factorized because of the average process. The free-space Green's tensor is given by
\begin{align}\label{eq:free-space_green}
   \bm{G}_0 (\bm{q}) &= \left[ (k_0^2 - q^2) \bm{I} + \bm{q} \otimes \bm{q}\right]^{-1}
\\\nonumber
   & = \left[ k_0^2 \bm{I} - q^2 \bm{P}(\hat{\bm{q}}) \right]^{-1},
\end{align}
with $\bm{P}(\hat{\bm{q}}) = \bm{I} - \hat{\bm{q}} \otimes \hat{\bm{q}}$. The average Green's tensor then reads
\begin{align}\label{eq:average_green}
   \bra \bm{G} (\bm{q}) \ket &= \left[ \bm{I} - \bm{G}_0 (\bm{q}) \bm{\Sigma}(\bm{q}) \right]^{-1} \bm{G}_0 (\bm{q})
\\\nonumber
   & = \left[ k_0^2 \bm{I} - q^2 \bm{P}(\hat{\bm{q}}) - \bm{\Sigma}(\bm{q})\right]^{-1}.
\end{align}
By identification between Eq.~(\ref{eq:free-space_green}) and Eq.~(\ref{eq:average_green}), one can define an
\textit{effective} wavevector $\bm{q}_\text{eff} = k_0^2 \bm{\epsilon}_\text{eff}(\bm{q})$, where
$\bm{\epsilon}_\text{eff}(\bm{q}) = \bm{I} - \bm{\Sigma}(\bm{q})/k_0^2$ is the effective medium permittivity tensor,
yielding
\begin{equation}\label{eq:average_green_2}
   \bra \bm{G} (\bm{q}) \ket = \left[\bm{q}_\text{eff}^2 - q^2 \bm{P}(\hat{\bm{q}})\right]^{-1}.
\end{equation}

In a dilute (3D) medium, interferences between successive scattering events can be neglected, and the self-energy can be
calculated keeping only the first term of the multiple-scattering expansion
\begin{align}\label{eq:self-energy}
   \bm{\Sigma}(\bm{r},\bm{r}') &\simeq k_0^4 \bra \delta\epsilon(\bm{r}) \bm{G}_0 (\bm{r}-\bm{r}') \delta\epsilon(\bm{r}') \ket, \nonumber \\
   &= u k_0^4 f(\bm{r}-\bm{r}') \bm{G}_0 (\bm{r}-\bm{r}'),
\end{align}
or in reciprocal space
\begin{equation}\label{eq:sigma}
   \bm{\Sigma}(\bm{q}) = u k_0^4 \int f(\bm{q}-\bm{q}') \bm{G}_0 (\bm{q}') \frac{d\bm{q}'}{8\pi^3}.
\end{equation}
For a delta-correlated disorder, $f(\bm{q}-\bm{q}') = 1$, we have
\begin{equation}
   \im \bm{\Sigma}(\bm{q}) = -u k_0^4\frac{q}{6 \pi}  \bm{I}.
\end{equation}
The real part of $\bm{\Sigma}$, which is typically very small for dilute media, is scalar as well. The effective medium
permittivity then becomes a scalar quantity:
\begin{equation}
   \bm{\epsilon}_\text{eff} \simeq 1 - \frac{\re\Sigma(q)}{k_0^2} - i\frac{u k_0^2 q}{6 \pi}.
\end{equation}
This allows rewriting Eq~(\ref{eq:average_green}), after some algebra, as~\cite{Tai1993}
\begin{equation}\label{eq:effective_green}
   \bra \bm{G} (\bm{q}) \ket = \frac{1}{k_0^2 \epsilon_\text{eff} - q^2}  \left[ \bm{I} - \frac{\bm{q} \otimes \bm{q}}{k_0^2 \epsilon_\text{eff}} \right],
\end{equation}
which is equivalent to Eq.~(\ref{eq:Green-mulet}).

The coherent (ballistic) intensity in a disordered medium $I_\text{coh}= |\bra \bm{E} \ket|^2$ decays exponentially
following the Beer-Lambert law
\begin{align}
   I_\text{coh} (z) &= I_\text{coh} (0) \exp \left[-2 k_0 \im(n_\text{eff})z \right], \nonumber \\
   &= I_\text{coh} (0) \exp \left[-z/\ell \right],
\end{align}
with $\ell=(2 k_0 \im[n_\text{eff}])^{-1}$ the extinction length, $n_\text{eff} = \sqrt{\epsilon_\text{eff}}$ the
effective refractive index and $z$ the propagation direction. Since $\im\epsilon_\text{eff} \ll \re\epsilon_\text{eff}$
(\ie $-\im\Sigma \ll k_0^2$), we have $\im(n_\text{eff}) \simeq -\im\Sigma/(2k_0q)$, thereby leading to
\begin{equation}\label{eq:im_sigma}
   \im\Sigma(\bm{q}) = -\frac{q}{\ell}, \qquad u = \frac{6\pi}{k_0^4 \ell}.
\end{equation}
For an arbitrary (non-delta) correlated disorder $f(\bm{q}-\bm{q}') \neq 1$, Eq.~(\ref{eq:self-energy}) indicates that
$\bm{\Sigma}$ should not be a scalar. Thus, the average Green's tensor in Eq.~(\ref{eq:average_green}) cannot possibly
take the form of Eq.~(\ref{eq:effective_green}); and Eq.~(\ref{eq:im_sigma}), which introduces the mean free path $\ell$
in the RTE [Eq.~(\ref{eq:RTE})], should be corrected. Our results are therefore expected to be strictly valid only for
short-range structural correlations (close to a delta-correlated potential), \ie for scattering anisotropy factors $g$ close to 0.

\section{Short-range correlation approximation}\label{sec:A2}

\begin{figure}
   \centering
	\includegraphics[width=\linewidth]{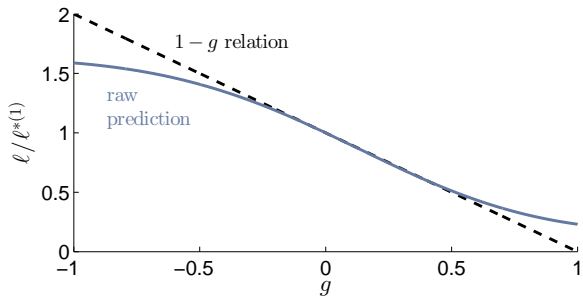}
   \caption{(Color online only) Evolution of the transport coefficient of
   the scalar mode, $1/\ell^{*(1)}$, in units of $1/\ell$, with the scattering anisotropy factor $g$. We compare the raw
   prediction obtained from the eigenmode decomposition, Eq.~(\ref{eq:eigenmode_decomposition}), before the development
   to order $g$ (solid cyan curve), with the expected $1-g$ scaling relation (black dashed curve). Our theoretical
   predictions are expected to be valid for short-range structural correlations, \ie for $g$ close to zero.}
   \label{fig2}
\end{figure}

As explained above, due to the fact that the self-energy $\bm{\Sigma}$ is assumed to be a scalar quantity in our model,
our theoretical predictions are expected to be valid only for short-range structural correlations, \ie for $g$ close to
zero. The validity range of this approximation can be apprehended by comparing the raw prediction obtained from  the
eigenmode decomposition, Eq.~(\ref{eq:eigenmode_decomposition}), \textit{without performing the development to order} $g$, with predictions from scalar theory. The eigenmode decomposition for the scalar mode, in the diffusion
approximation (\ie, to order $K^2$), yields a transport mean free path
\begin{equation}\label{eq:ltscalar}
   \ell^{*(1)} = \frac{10 + g(5g-2)}{10 + g(5g-12)} \ell,
\end{equation}
to be compared with the expected relation, $\ell^*=\ell(1-g)^{-1}$. The two relations are shown in Fig.~\ref{fig2},
where it is found that our prediction remains fairly good for $-0.3 \lesssim g \lesssim 0.3$, hence the range chosen in
Fig.~\ref{fig1}. Developing the transport coefficient $1/\ell^{*(1)}$ of Eq.~(\ref{eq:ltscalar}) to order $g$ yields the
proper $(1-g)$ scaling, as reported in Table~\ref{tab:diffcorr}.

\section{Eigenmode decomposition for uncorrelated disorder}\label{sec:A3}

For uncorrelated disorder, the scattering anisotropy factor $g$ equals zero, such that, from
Eq.~(\ref{eq:first_moment}), we immediately obtain
\begin{equation}
   i \bm{K} \cdot \bm{j}_{ijkl}(\bm{K}) = \frac{4\pi}{3} K^2 \ell L^{(0)}_{ijkl}(\bm{K}).
\end{equation}
Inserting it into Eq.~(\ref{eq:zeroth_moment}), we get
\begin{align}
   \frac{4\pi}{3} & K^2 \ell L^{(0)}_{ijkl}(\bm{K}) + \frac{4\pi}{\ell} L^{(0)}_{ijkl}(\bm{K}) \nonumber \\
   & = \frac{2}{3}S_{ijkl} + \frac{4\pi}{\ell} S_{ijmn} L^{(0)}_{mnkl}(\bm{K}).
\end{align}
Performing an eigenmode decomposition of $S_{ijkl}$ as
\begin{equation}
   S_{ijkl}=\sum_{p=1}^9 S^{(p)} \left|ij\ket^{(p)} \bra kl \right|^{(p)},
\end{equation}
and similarly for $L^{(0)}_{ijkl}= c/(6\pi) U_{ijkl}$, we directly find that the diffusion of the energy density
in each polarization eigenchannel, $U^{(p)}$, follows the solution of the diffusion equation,
Eq.~(\ref{eq:diffusionsolution}), with
\begin{equation}
   \mathcal{D}^{(p)} = \frac{c\ell}{3} \frac{1}{S^{(p)}}, \qquad \mu_a^{(p)} = \frac{1}{\ell} \frac{1-S^{(p)}}{S^{(p)}}.
\end{equation}
The eigenvalues of $S_{ijkl}$ are $1$, $1/2$ and $7/10$ with degeneracies $1$, $3$ and $5$, respectively, thereby leading to the
values reported in Table~\ref{tab:diffuncorr}.


\begin{thebibliography}{49}
\expandafter\ifx\csname natexlab\endcsname\relax\def\natexlab#1{#1}\fi
\expandafter\ifx\csname bibnamefont\endcsname\relax
  \def\bibnamefont#1{#1}\fi
\expandafter\ifx\csname bibfnamefont\endcsname\relax
  \def\bibfnamefont#1{#1}\fi
\expandafter\ifx\csname citenamefont\endcsname\relax
  \def\citenamefont#1{#1}\fi
\expandafter\ifx\csname url\endcsname\relax
  \def\url#1{\texttt{#1}}\fi
\expandafter\ifx\csname urlprefix\endcsname\relax\def\urlprefix{URL }\fi
\providecommand{\bibinfo}[2]{#2}
\providecommand{\eprint}[2][]{\url{#2}}

\bibitem[{\citenamefont{Akkermans and Montambaux}(2011)}]{Akkermans2011}
\bibinfo{author}{\bibfnamefont{E.}~\bibnamefont{Akkermans}} \bibnamefont{and}
  \bibinfo{author}{\bibfnamefont{G.}~\bibnamefont{Montambaux}},
  \emph{\bibinfo{title}{{Mesoscopic Physics of Electrons and Photons}}}
  (\bibinfo{publisher}{Cambridge University Press}, \bibinfo{year}{2011}).

\bibitem[{\citenamefont{Sheng}(2010)}]{Sheng2010}
\bibinfo{author}{\bibfnamefont{P.}~\bibnamefont{Sheng}},
  \emph{\bibinfo{title}{{Introduction to Wave Scattering, Localization and
  Mesoscopic Phenomena}}} (\bibinfo{publisher}{Springer},
  \bibinfo{year}{2010}).

\bibitem[{\citenamefont{Goodman}(2015)}]{Goodman2015}
\bibinfo{author}{\bibfnamefont{J.~W.} \bibnamefont{Goodman}},
  \emph{\bibinfo{title}{{Statistical Optics}}} (\bibinfo{publisher}{Wiley},
  \bibinfo{year}{2015}).

\bibitem[{\citenamefont{Brosseau}(1998)}]{Brosseau1998}
\bibinfo{author}{\bibfnamefont{C.}~\bibnamefont{Brosseau}},
  \emph{\bibinfo{title}{{Fundamentals of Polarized Light: A Statistical Optics
  Approach}}} (\bibinfo{publisher}{Wiley-Blackwell}, \bibinfo{year}{1998}).

\bibitem[{\citenamefont{Tuchin et~al.}(2006)\citenamefont{Tuchin, Wang, and
  Zimnyakov}}]{Tuchin2006}
\bibinfo{author}{\bibfnamefont{V.~V.} \bibnamefont{Tuchin}},
  \bibinfo{author}{\bibfnamefont{L.}~\bibnamefont{Wang}}, \bibnamefont{and}
  \bibinfo{author}{\bibfnamefont{D.~A.} \bibnamefont{Zimnyakov}},
  \emph{\bibinfo{title}{{Optical Polarization in Biomedical Applications}}} (\bibinfo{publisher}{Springer},
  \bibinfo{year}{2006}).

\bibitem[{\citenamefont{Maradudin}(2007)}]{Maradudin2007}
\bibinfo{editor}{\bibfnamefont{A.~A.} \bibnamefont{Maradudin}}, ed.,
  \emph{\bibinfo{title}{{Light Scattering and Nanoscale Surface Roughness}}}
  (\bibinfo{publisher}{Springer}, \bibinfo{year}{2007}).

\bibitem[{\citenamefont{Andrews and Phillips}(2005)}]{Andrews2005}
\bibinfo{author}{\bibfnamefont{L.~C.} \bibnamefont{Andrews}} \bibnamefont{and}
  \bibinfo{author}{\bibfnamefont{R.~L.} \bibnamefont{Phillips}},
  \emph{\bibinfo{title}{{Laser Beam Propagation through Random Media}}}
  (\bibinfo{publisher}{SPIE}, \bibinfo{year}{2005}).

\bibitem[{\citenamefont{Shirai et~al.}(2003)\citenamefont{Shirai, Dogariu, and
  Wolf}}]{Shirai2003}
\bibinfo{author}{\bibfnamefont{T.}~\bibnamefont{Shirai}},
  \bibinfo{author}{\bibfnamefont{A.}~\bibnamefont{Dogariu}}, \bibnamefont{and}
  \bibinfo{author}{\bibfnamefont{E.}~\bibnamefont{Wolf}}, \bibinfo{journal}{J.
  Opt. Soc. Am. A} \textbf{\bibinfo{volume}{20}}, \bibinfo{pages}{1094}
  (\bibinfo{year}{2003}).

\bibitem[{\citenamefont{Apostol and Dogariu}(2003)}]{Apostol2003}
\bibinfo{author}{\bibfnamefont{A.}~\bibnamefont{Apostol}} \bibnamefont{and}
  \bibinfo{author}{\bibfnamefont{A.}~\bibnamefont{Dogariu}},
  \bibinfo{journal}{Phys. Rev. Lett.} \textbf{\bibinfo{volume}{91}},
  \bibinfo{pages}{093901} (\bibinfo{year}{2003}).

\bibitem[{\citenamefont{Carminati}(2010)}]{Carminati2010}
\bibinfo{author}{\bibfnamefont{R.}~\bibnamefont{Carminati}},
  \bibinfo{journal}{Phys. Rev. A} \textbf{\bibinfo{volume}{81}},
  \bibinfo{pages}{1} (\bibinfo{year}{2010}).

\bibitem[{\citenamefont{Parigi et~al.}(2016)\citenamefont{Parigi, Perros,
  Binard, Bourdillon, Ma{\^{i}}tre, Carminati, Krachmalnicoff, and {De
  Wilde}}}]{Parigi2016}
\bibinfo{author}{\bibfnamefont{V.}~\bibnamefont{Parigi}},
  \bibinfo{author}{\bibfnamefont{E.}~\bibnamefont{Perros}},
  \bibinfo{author}{\bibfnamefont{G.}~\bibnamefont{Binard}},
  \bibinfo{author}{\bibfnamefont{C.}~\bibnamefont{Bourdillon}},
  \bibinfo{author}{\bibfnamefont{A.}~\bibnamefont{Ma{\^{i}}tre}},
  \bibinfo{author}{\bibfnamefont{R.}~\bibnamefont{Carminati}},
  \bibinfo{author}{\bibfnamefont{V.}~\bibnamefont{Krachmalnicoff}},
  \bibnamefont{and} \bibinfo{author}{\bibfnamefont{Y.}~\bibnamefont{{De
  Wilde}}}, \bibinfo{journal}{Opt. Express} \textbf{\bibinfo{volume}{24}},
  \bibinfo{pages}{7019} (\bibinfo{year}{2016}).

\bibitem[{\citenamefont{Caz{\'{e}} et~al.}(2010)\citenamefont{Caz{\'{e}},
  Pierrat, and Carminati}}]{Caze2010}
\bibinfo{author}{\bibfnamefont{A.}~\bibnamefont{Caz{\'{e}}}},
  \bibinfo{author}{\bibfnamefont{R.}~\bibnamefont{Pierrat}}, \bibnamefont{and}
  \bibinfo{author}{\bibfnamefont{R.}~\bibnamefont{Carminati}},
  \bibinfo{journal}{Phys. Rev. A} \textbf{\bibinfo{volume}{82}},
  \bibinfo{pages}{043823} (\bibinfo{year}{2010}).

\bibitem[{\citenamefont{Sapienza et~al.}(2011)\citenamefont{Sapienza,
  Bondareff, Pierrat, Habert, Carminati, and {Van Hulst}}}]{Sapienza2011a}
\bibinfo{author}{\bibfnamefont{R.}~\bibnamefont{Sapienza}},
  \bibinfo{author}{\bibfnamefont{P.}~\bibnamefont{Bondareff}},
  \bibinfo{author}{\bibfnamefont{R.}~\bibnamefont{Pierrat}},
  \bibinfo{author}{\bibfnamefont{B.}~\bibnamefont{Habert}},
  \bibinfo{author}{\bibfnamefont{R.}~\bibnamefont{Carminati}},
  \bibnamefont{and} \bibinfo{author}{\bibfnamefont{N.~F.} \bibnamefont{{Van
  Hulst}}}, \bibinfo{journal}{Phys. Rev. Lett.} \textbf{\bibinfo{volume}{106}},
  \bibinfo{pages}{1} (\bibinfo{year}{2011}).

\bibitem[{\citenamefont{Schmidt et~al.}(2015)\citenamefont{Schmidt, Aizpurua,
  Zambrana-Puyalto, Vidal, Molina-Terriza, and S{\'{a}}enz}}]{Schmidt2015}
\bibinfo{author}{\bibfnamefont{M.~K.} \bibnamefont{Schmidt}},
  \bibinfo{author}{\bibfnamefont{J.}~\bibnamefont{Aizpurua}},
  \bibinfo{author}{\bibfnamefont{X.}~\bibnamefont{Zambrana-Puyalto}},
  \bibinfo{author}{\bibfnamefont{X.}~\bibnamefont{Vidal}},
  \bibinfo{author}{\bibfnamefont{G.}~\bibnamefont{Molina-Terriza}},
  \bibnamefont{and} \bibinfo{author}{\bibfnamefont{J.~J.}
  \bibnamefont{S{\'{a}}enz}}, \bibinfo{journal}{Phys. Rev. Lett.}
  \textbf{\bibinfo{volume}{114}}, \bibinfo{pages}{113902}
  (\bibinfo{year}{2015}).

\bibitem[{\citenamefont{Torquato}(2005)}]{Torquato2005}
\bibinfo{author}{\bibfnamefont{S.}~\bibnamefont{Torquato}},
  \emph{\bibinfo{title}{{Random Heterogeneous Materials: Microstructure and
  Macroscopic Properties}}} (\bibinfo{publisher}{Springer},
  \bibinfo{year}{2005}).

\bibitem[{\citenamefont{Rojas-Ochoa
  et~al.}(2004{\natexlab{a}})\citenamefont{Rojas-Ochoa, Mendez-Alcaraz,
  S{\'{a}}enz, Schurtenberger, and Scheffold}}]{RojasOchoa2004}
\bibinfo{author}{\bibfnamefont{L.~F.} \bibnamefont{Rojas-Ochoa}},
  \bibinfo{author}{\bibfnamefont{J.~M.} \bibnamefont{Mendez-Alcaraz}},
  \bibinfo{author}{\bibfnamefont{J.~J.} \bibnamefont{S{\'{a}}enz}},
  \bibinfo{author}{\bibfnamefont{P.}~\bibnamefont{Schurtenberger}},
  \bibnamefont{and}
  \bibinfo{author}{\bibfnamefont{F.}~\bibnamefont{Scheffold}},
  \bibinfo{journal}{Phys. Rev. Lett.} \textbf{\bibinfo{volume}{93}},
  \bibinfo{pages}{73903} (\bibinfo{year}{2004}{\natexlab{a}}).

\bibitem[{\citenamefont{Garc{\'{i}}a et~al.}(2007)\citenamefont{Garc{\'{i}}a,
  Sapienza, Blanco, and L{\'{o}}pez}}]{Garcia2007}
\bibinfo{author}{\bibfnamefont{P.}~\bibnamefont{Garc{\'{i}}a}},
  \bibinfo{author}{\bibfnamefont{R.}~\bibnamefont{Sapienza}},
  \bibinfo{author}{\bibfnamefont{{\'{A}}.}~\bibnamefont{Blanco}},
  \bibnamefont{and}
  \bibinfo{author}{\bibfnamefont{C.}~\bibnamefont{L{\'{o}}pez}},
  \bibinfo{journal}{Adv. Mater.} \textbf{\bibinfo{volume}{19}},
  \bibinfo{pages}{2597} (\bibinfo{year}{2007}).

\bibitem[{\citenamefont{MacKintosh et~al.}(1989)\citenamefont{MacKintosh, Zhu,
  Pine, and Weitz}}]{MacKintosh1989a}
\bibinfo{author}{\bibfnamefont{F.~C.} \bibnamefont{MacKintosh}},
  \bibinfo{author}{\bibfnamefont{J.~X.} \bibnamefont{Zhu}},
  \bibinfo{author}{\bibfnamefont{D.~J.} \bibnamefont{Pine}}, \bibnamefont{and}
  \bibinfo{author}{\bibfnamefont{D.~A.} \bibnamefont{Weitz}},
  \bibinfo{journal}{Phys. Rev. B} \textbf{\bibinfo{volume}{40}},
  \bibinfo{pages}{9342} (\bibinfo{year}{1989}).

\bibitem[{\citenamefont{Xu and Alfano}(2005{\natexlab{a}})}]{Xu2005a}
\bibinfo{author}{\bibfnamefont{M.}~\bibnamefont{Xu}} \bibnamefont{and}
  \bibinfo{author}{\bibfnamefont{R.~R.} \bibnamefont{Alfano}},
  \bibinfo{journal}{Phys. Rev. E. Stat. Nonlin. Soft Matter Phys.}
  \textbf{\bibinfo{volume}{72}}, \bibinfo{pages}{065601}
  (\bibinfo{year}{2005}{\natexlab{a}}).

\bibitem[{\citenamefont{Gorodnichev et~al.}(2007)\citenamefont{Gorodnichev,
  Kuzovlev, and Rogozkin}}]{Gorodnichev2007}
\bibinfo{author}{\bibfnamefont{E.~E.} \bibnamefont{Gorodnichev}},
  \bibinfo{author}{\bibfnamefont{A.~I.} \bibnamefont{Kuzovlev}},
  \bibnamefont{and} \bibinfo{author}{\bibfnamefont{D.~B.}
  \bibnamefont{Rogozkin}}, \bibinfo{journal}{J. Exp. Theor. Phys.}
  \textbf{\bibinfo{volume}{104}}, \bibinfo{pages}{319} (\bibinfo{year}{2007}).

\bibitem[{\citenamefont{Skipetrov and Sokolov}(2014)}]{Skipetrov2014}
\bibinfo{author}{\bibfnamefont{S.~E.} \bibnamefont{Skipetrov}}
  \bibnamefont{and} \bibinfo{author}{\bibfnamefont{I.~M.}
  \bibnamefont{Sokolov}}, \bibinfo{journal}{Phys. Rev. Lett.}
  \textbf{\bibinfo{volume}{112}}, \bibinfo{pages}{23905}
  (\bibinfo{year}{2014}).

\bibitem[{\citenamefont{Bellando et~al.}(2014)\citenamefont{Bellando, Gero,
  Akkermans, and Kaiser}}]{Bellando2014}
\bibinfo{author}{\bibfnamefont{L.}~\bibnamefont{Bellando}},
  \bibinfo{author}{\bibfnamefont{A.}~\bibnamefont{Gero}},
  \bibinfo{author}{\bibfnamefont{E.}~\bibnamefont{Akkermans}},
  \bibnamefont{and} \bibinfo{author}{\bibfnamefont{R.}~\bibnamefont{Kaiser}},
  \bibinfo{journal}{Phys. Rev. A} \textbf{\bibinfo{volume}{90}},
  \bibinfo{pages}{063822} (\bibinfo{year}{2014}).

\bibitem[{\citenamefont{Edagawa et~al.}(2008)\citenamefont{Edagawa, Kanoko, and
  Notomi}}]{Edagawa2008}
\bibinfo{author}{\bibfnamefont{K.}~\bibnamefont{Edagawa}},
  \bibinfo{author}{\bibfnamefont{S.}~\bibnamefont{Kanoko}}, \bibnamefont{and}
  \bibinfo{author}{\bibfnamefont{M.}~\bibnamefont{Notomi}},
  \bibinfo{journal}{Phys. Rev. Lett.} \textbf{\bibinfo{volume}{100}},
  \bibinfo{pages}{1} (\bibinfo{year}{2008}).

\bibitem[{\citenamefont{Liew et~al.}(2011)\citenamefont{Liew, Yang, Noh,
  Schreck, Dufresne, O'Hern, and Cao}}]{Liew2011}
\bibinfo{author}{\bibfnamefont{S.~F.} \bibnamefont{Liew}},
  \bibinfo{author}{\bibfnamefont{J.-K.} \bibnamefont{Yang}},
  \bibinfo{author}{\bibfnamefont{H.}~\bibnamefont{Noh}},
  \bibinfo{author}{\bibfnamefont{C.~F.} \bibnamefont{Schreck}},
  \bibinfo{author}{\bibfnamefont{E.~R.} \bibnamefont{Dufresne}},
  \bibinfo{author}{\bibfnamefont{C.~S.} \bibnamefont{O'Hern}},
  \bibnamefont{and} \bibinfo{author}{\bibfnamefont{H.}~\bibnamefont{Cao}},
  \bibinfo{journal}{Phys. Rev. A} \textbf{\bibinfo{volume}{84}},
  \bibinfo{pages}{63818} (\bibinfo{year}{2011}).

\bibitem[{\citenamefont{Imagawa et~al.}(2010)\citenamefont{Imagawa, Edagawa,
  Morita, Niino, Kagawa, and Notomi}}]{Imagawa2010}
\bibinfo{author}{\bibfnamefont{S.}~\bibnamefont{Imagawa}},
  \bibinfo{author}{\bibfnamefont{K.}~\bibnamefont{Edagawa}},
  \bibinfo{author}{\bibfnamefont{K.}~\bibnamefont{Morita}},
  \bibinfo{author}{\bibfnamefont{T.}~\bibnamefont{Niino}},
  \bibinfo{author}{\bibfnamefont{Y.}~\bibnamefont{Kagawa}}, \bibnamefont{and}
  \bibinfo{author}{\bibfnamefont{M.}~\bibnamefont{Notomi}},
  \bibinfo{journal}{Phys. Rev. B} \textbf{\bibinfo{volume}{82}},
  \bibinfo{pages}{115116} (\bibinfo{year}{2010}).

\bibitem[{\citenamefont{Chandrasekhar}(1960)}]{Chandrasekhar1960}
\bibinfo{author}{\bibfnamefont{S.}~\bibnamefont{Chandrasekhar}},
  \emph{\bibinfo{title}{{Radiative Transfer}}} (\bibinfo{publisher}{Dover
  Publications}, \bibinfo{year}{1960}).

\bibitem[{\citenamefont{Papanicolaou and Burridge}(1975)}]{Papanicolaou1975}
\bibinfo{author}{\bibfnamefont{G.~C.} \bibnamefont{Papanicolaou}}
  \bibnamefont{and} \bibinfo{author}{\bibfnamefont{R.}~\bibnamefont{Burridge}},
  \bibinfo{journal}{J. Math. Phys.} \textbf{\bibinfo{volume}{16}},
  \bibinfo{pages}{2074} (\bibinfo{year}{1975}).

\bibitem[{\citenamefont{Mishchenko and Travis}(2006)}]{Mishchenko2006}
\bibinfo{author}{\bibfnamefont{M.~I.} \bibnamefont{Mishchenko}}
  \bibnamefont{and} \bibinfo{author}{\bibfnamefont{L.~D.}
  \bibnamefont{Travis}}, \emph{\bibinfo{title}{{Multiple Scattering of Light by
  Particles: Radiative Transfer and Coherent Backscattering}}}
  (\bibinfo{publisher}{Cambridge University Press}, \bibinfo{year}{2006}).

\bibitem[{\citenamefont{Amic et~al.}(1997)\citenamefont{Amic, Luck, and
  Nieuwenhuizen}}]{Amic1997}
\bibinfo{author}{\bibfnamefont{E.}~\bibnamefont{Amic}},
  \bibinfo{author}{\bibfnamefont{J.~M.} \bibnamefont{Luck}}, \bibnamefont{and}
  \bibinfo{author}{\bibfnamefont{T.~M.} \bibnamefont{Nieuwenhuizen}},
  \bibinfo{journal}{J. Phys. I} \textbf{\bibinfo{volume}{7}},
  \bibinfo{pages}{445} (\bibinfo{year}{1997}).

\bibitem[{\citenamefont{Gorodnichev et~al.}(2014)\citenamefont{Gorodnichev,
  Kuzovlev, and Rogozkin}}]{Gorodnichev2014}
\bibinfo{author}{\bibfnamefont{E.~E.} \bibnamefont{Gorodnichev}},
  \bibinfo{author}{\bibfnamefont{A.~I.} \bibnamefont{Kuzovlev}},
  \bibnamefont{and} \bibinfo{author}{\bibfnamefont{D.~B.}
  \bibnamefont{Rogozkin}}, \bibinfo{journal}{Phys. Rev. E}
  \textbf{\bibinfo{volume}{90}}, \bibinfo{pages}{043205}
  (\bibinfo{year}{2014}).

\bibitem[{\citenamefont{Akkermans et~al.}(1988)\citenamefont{Akkermans, Wolf,
  Maynard, and Maret}}]{Akkermans1988}
\bibinfo{author}{\bibfnamefont{E.}~\bibnamefont{Akkermans}},
  \bibinfo{author}{\bibfnamefont{P.}~\bibnamefont{Wolf}},
  \bibinfo{author}{\bibfnamefont{R.}~\bibnamefont{Maynard}}, \bibnamefont{and}
  \bibinfo{author}{\bibfnamefont{G.}~\bibnamefont{Maret}}, \bibinfo{journal}{J.
  Phys.} \textbf{\bibinfo{volume}{49}}, \bibinfo{pages}{77}
  (\bibinfo{year}{1988}).

\bibitem[{\citenamefont{Xu and Alfano}(2005{\natexlab{b}})}]{Xu2005}
\bibinfo{author}{\bibfnamefont{M.}~\bibnamefont{Xu}} \bibnamefont{and}
  \bibinfo{author}{\bibfnamefont{R.~R.} \bibnamefont{Alfano}},
  \bibinfo{journal}{Phys. Rev. Lett.} \textbf{\bibinfo{volume}{95}},
  \bibinfo{pages}{213901} (\bibinfo{year}{2005}{\natexlab{b}}).

\bibitem[{\citenamefont{Rojas-Ochoa
  et~al.}(2004{\natexlab{b}})\citenamefont{Rojas-Ochoa, Lacoste, Lenke,
  Schurtenberger, and Scheffold}}]{Rojas-Ochoa2004a}
\bibinfo{author}{\bibfnamefont{L.~F.} \bibnamefont{Rojas-Ochoa}},
  \bibinfo{author}{\bibfnamefont{D.}~\bibnamefont{Lacoste}},
  \bibinfo{author}{\bibfnamefont{R.}~\bibnamefont{Lenke}},
  \bibinfo{author}{\bibfnamefont{P.}~\bibnamefont{Schurtenberger}},
  \bibnamefont{and}
  \bibinfo{author}{\bibfnamefont{F.}~\bibnamefont{Scheffold}},
  \bibinfo{journal}{J. Opt. Soc. Am. A. Opt. Image Sci. Vis.}
  \textbf{\bibinfo{volume}{21}}, \bibinfo{pages}{1799}
  (\bibinfo{year}{2004}{\natexlab{b}}).

\bibitem[{\citenamefont{Stephen and Cwilich}(1986)}]{Stephen1986}
\bibinfo{author}{\bibfnamefont{M.~J.} \bibnamefont{Stephen}} \bibnamefont{and}
  \bibinfo{author}{\bibfnamefont{G.}~\bibnamefont{Cwilich}},
  \bibinfo{journal}{Phys. Rev. B} \textbf{\bibinfo{volume}{34}},
  \bibinfo{pages}{7564} (\bibinfo{year}{1986}).

\bibitem[{\citenamefont{MacKintosh and John}(1988)}]{MacKintosh1988}
\bibinfo{author}{\bibfnamefont{F.~C.} \bibnamefont{MacKintosh}}
  \bibnamefont{and} \bibinfo{author}{\bibfnamefont{S.}~\bibnamefont{John}},
  \bibinfo{journal}{Phys. Rev. B} \textbf{\bibinfo{volume}{37}},
  \bibinfo{pages}{1884} (\bibinfo{year}{1988}).

\bibitem[{\citenamefont{Ozrin}(1992)}]{Ozrin1992}
\bibinfo{author}{\bibfnamefont{V.~D.} \bibnamefont{Ozrin}},
  \bibinfo{journal}{Waves in Random Media} \textbf{\bibinfo{volume}{2}},
  \bibinfo{pages}{141} (\bibinfo{year}{1992}).

\bibitem[{\citenamefont{van Tiggelen et~al.}(1996)\citenamefont{van Tiggelen,
  Maynard, and Nieuwenhuizen}}]{VanTiggelen1996}
\bibinfo{author}{\bibfnamefont{B.}~\bibnamefont{van Tiggelen}},
  \bibinfo{author}{\bibfnamefont{R.}~\bibnamefont{Maynard}}, \bibnamefont{and}
  \bibinfo{author}{\bibfnamefont{T.}~\bibnamefont{Nieuwenhuizen}},
  \bibinfo{journal}{Phys. Rev. E} \textbf{\bibinfo{volume}{53}},
  \bibinfo{pages}{2881} (\bibinfo{year}{1996}).

\bibitem[{\citenamefont{{van Tiggelen} et~al.}(1999)\citenamefont{{van
  Tiggelen}, Lagendijk, and Tip}}]{VanTiggelen1999}
\bibinfo{author}{\bibfnamefont{B.~A.} \bibnamefont{{van Tiggelen}}},
  \bibinfo{author}{\bibfnamefont{A.}~\bibnamefont{Lagendijk}},
  \bibnamefont{and} \bibinfo{author}{\bibfnamefont{A.}~\bibnamefont{Tip}},
  \bibinfo{journal}{J. Phys. Condens. Matter} \textbf{\bibinfo{volume}{2}},
  \bibinfo{pages}{7653} (\bibinfo{year}{1999}).

\bibitem[{\citenamefont{M{\"{u}}ller and Miniatura}(2002)}]{Muller2002}
\bibinfo{author}{\bibfnamefont{C.~A.} \bibnamefont{M{\"{u}}ller}}
  \bibnamefont{and}
  \bibinfo{author}{\bibfnamefont{C.}~\bibnamefont{Miniatura}},
  \bibinfo{journal}{J. Phys. A Math. Gen.} \textbf{\bibinfo{volume}{35}},
  \bibinfo{pages}{10163} (\bibinfo{year}{2002}).

\bibitem[{\citenamefont{Vynck et~al.}(2014)\citenamefont{Vynck, Pierrat, and
  Carminati}}]{Vynck2014}
\bibinfo{author}{\bibfnamefont{K.}~\bibnamefont{Vynck}},
  \bibinfo{author}{\bibfnamefont{R.}~\bibnamefont{Pierrat}}, \bibnamefont{and}
  \bibinfo{author}{\bibfnamefont{R.}~\bibnamefont{Carminati}},
  \bibinfo{journal}{Phys. Rev. A} \textbf{\bibinfo{volume}{89}},
  \bibinfo{pages}{013842} (\bibinfo{year}{2014}).

\bibitem[{\citenamefont{Set{\"{a}}l{\"{a}}
  et~al.}(2002)\citenamefont{Set{\"{a}}l{\"{a}}, Shevchenko, Kaivola, and
  Friberg}}]{Setala2002}
\bibinfo{author}{\bibfnamefont{T.}~\bibnamefont{Set{\"{a}}l{\"{a}}}},
  \bibinfo{author}{\bibfnamefont{A.}~\bibnamefont{Shevchenko}},
  \bibinfo{author}{\bibfnamefont{M.}~\bibnamefont{Kaivola}}, \bibnamefont{and}
  \bibinfo{author}{\bibfnamefont{A.~T.} \bibnamefont{Friberg}},
  \bibinfo{journal}{Phys. Rev. E} \textbf{\bibinfo{volume}{66}},
  \bibinfo{pages}{1} (\bibinfo{year}{2002}).

\bibitem[{\citenamefont{Dennis}(2007)}]{Dennis2007}
\bibinfo{author}{\bibfnamefont{M.~R.} \bibnamefont{Dennis}},
  \bibinfo{journal}{J. Opt. Soc. Am. A} \textbf{\bibinfo{volume}{24}},
  \bibinfo{pages}{2065} (\bibinfo{year}{2007}).

\bibitem[{\citenamefont{R{\'{e}}fr{\'{e}}gier
  et~al.}(2014)\citenamefont{R{\'{e}}fr{\'{e}}gier, Wasik, Vynck, and
  Carminati}}]{Refregier2014}
\bibinfo{author}{\bibfnamefont{P.}~\bibnamefont{R{\'{e}}fr{\'{e}}gier}},
  \bibinfo{author}{\bibfnamefont{V.}~\bibnamefont{Wasik}},
  \bibinfo{author}{\bibfnamefont{K.}~\bibnamefont{Vynck}}, \bibnamefont{and}
  \bibinfo{author}{\bibfnamefont{R.}~\bibnamefont{Carminati}},
  \bibinfo{journal}{Opt. Lett.} \textbf{\bibinfo{volume}{39}},
  \bibinfo{pages}{2362} (\bibinfo{year}{2014}).

\bibitem[{\citenamefont{Gil}(2014)}]{Gil2014}
\bibinfo{author}{\bibfnamefont{J.~J.} \bibnamefont{Gil}},
  \bibinfo{journal}{Phys. Rev. A} \textbf{\bibinfo{volume}{90}},
  \bibinfo{pages}{043858} (\bibinfo{year}{2014}).

\bibitem[{\citenamefont{Dogariu and Carminati}(2015)}]{Dogariu2015}
\bibinfo{author}{\bibfnamefont{A.}~\bibnamefont{Dogariu}} \bibnamefont{and}
  \bibinfo{author}{\bibfnamefont{R.}~\bibnamefont{Carminati}},
  \bibinfo{journal}{Phys. Rep.} \textbf{\bibinfo{volume}{559}},
  \bibinfo{pages}{1} (\bibinfo{year}{2015}).

\bibitem[{\citenamefont{Tai}(1993)}]{Tai1993}
\bibinfo{author}{\bibfnamefont{C.}~\bibnamefont{Tai}},
  \emph{\bibinfo{title}{{Dyadic Green Functions in Electromagnetic Theory}}}
  (\bibinfo{publisher}{IEEE Press, New-York}, \bibinfo{year}{1993}).

\bibitem[{\citenamefont{Arnoldus}(2003)}]{Arnoldus2003}
\bibinfo{author}{\bibfnamefont{H.~F.} \bibnamefont{Arnoldus}},
  \bibinfo{journal}{J. Mod. Opt.} \textbf{\bibinfo{volume}{50}},
  \bibinfo{pages}{755} (\bibinfo{year}{2003}).

\bibitem[{Mat()}]{Mathematica}
\emph{\bibinfo{title}{{Wolfram Mathematica}}}.

\bibitem[{\citenamefont{Henyey and Greenstein}(1941)}]{Henyey1941}
\bibinfo{author}{\bibfnamefont{L.~C.} \bibnamefont{Henyey}} \bibnamefont{and}
  \bibinfo{author}{\bibfnamefont{J.~L.} \bibnamefont{Greenstein}},
  \bibinfo{journal}{Astrophys. J.} \textbf{\bibinfo{volume}{93}},
  \bibinfo{pages}{70} (\bibinfo{year}{1941}).

\end{thebibliography}

\end{document}